\documentclass[prl,a4paper,twocolumn,superscriptaddress,preprintnumbers]{revtex4-1}
\pdfoutput=1
\usepackage[utf8x]{inputenc}
\usepackage{enumitem}
\usepackage{amssymb}
\usepackage{amsmath}
\usepackage{graphicx}
\usepackage{bbm}
\usepackage{psfrag}
\usepackage{latexsym}
\usepackage[usenames, dvipsnames]{color}
\usepackage{xcolor}	
\usepackage[unicode,implicit]{hyperref}
\hypersetup{
    colorlinks,
    linkcolor={blue!50!black},
    citecolor={blue!50!black},
    urlcolor={blue!80!black}
}

\newcommand{\be}{\begin{equation}}
\newcommand{\ee}{\end{equation}}
\newcommand{\bea}{\begin{eqnarray}}
\newcommand{\eea}{\end{eqnarray}}
\newcommand{\ba}{\begin{eqnarray}}
\newcommand{\ea}{\end{eqnarray}}

\newcommand{\beq}{\begin{equation}}
\newcommand{\eeq}{\end{equation}}
\newcommand{\beqa}{\begin{eqnarray}}
\newcommand{\eeqa}{\end{eqnarray}}
\newcommand{\beqar}{\begin{eqnarray*}}
\newcommand{\eeqar}{\end{eqnarray*}}

\newcommand{\eq}{\begin{equation}}
\newcommand{\eqx}{\end{equation}}
\newcommand{\eqn}{\begin{eqnarray}}
\newcommand{\eqnx}{\end{eqnarray}}

\usepackage[normalem]{ulem}

\pdfoutput=1

\usepackage{latexsym,amsthm,amsfonts,amsmath,mathrsfs,amssymb}
\usepackage{booktabs} 

\hypersetup{%
  pdftitle    = {Regular Stringy Black Holes?}
  pdfkeywords = {black hole, singularity, string theory,
    branes, gravity, supergravity, superstrings, supersymmetry, corrections, higher-order},
  pdfauthor   = {Pablo A. Cano, Samuele Chimento, Tomas Ortin, Alejandro Ruiperez},
  plainpages  = true,
  colorlinks  = true,
  citecolor   = blue,
  urlcolor    = red,
  linkcolor   = black
}

\begin{document}


\author{Pablo A. Cano}\email{pablo.cano@uam.es}

\affiliation{\it Instituto de F\'isica Te\'orica UAM/CSIC, C/ Nicol\'as Cabrera, 13-15, C.U. Cantoblanco, 28049 Madrid, Spain}

\author{Samuele Chimento}\email{samuele.chimento@csic.es}

\affiliation{\it Instituto de F\'isica Te\'orica UAM/CSIC, C/ Nicol\'as Cabrera, 13-15, C.U. Cantoblanco, 28049 Madrid, Spain}

\author{Tom\'{a}s Ort\'{\i}n}\email{tomas.ortin@csic.es}

\affiliation{\it Instituto de F\'isica Te\'orica UAM/CSIC, C/ Nicol\'as Cabrera, 13-15, C.U. Cantoblanco, 28049 Madrid, Spain}

\author{Alejandro~Ruip\'erez}\email{alejandro.ruiperez@uam.es}

\affiliation{\it Instituto de F\'isica Te\'orica UAM/CSIC, C/ Nicol\'as Cabrera, 13-15, C.U. Cantoblanco, 28049 Madrid, Spain}

\title{Regular Stringy Black Holes?}

\date{June 21, 2018}
\preprint{IFT-UAM/CSIC-18-62}

\renewcommand{\thefootnote}{\alph{footnote}}
\setcounter{footnote}{0}
\renewcommand{\thefootnote}{\arabic{footnote}}

\begin{abstract}
We study the first-order $\alpha'$ corrections to the singular
    4-dimensional massless stringy black holes studied in the nineties in the
    context of the Heterotic Superstring. We show that the $\alpha'$ corrections not only induce
    a non-vanishing mass and give rise to an event horizon, but also eliminate
    the singularity giving rise to a regular spacetime whose global structure
    includes further asymptotically flat regions in which the spacetime's mass
    is positive or negative. We study the timelike and null geodesics
    and their effective potential, showing that the spacetime is geodesically complete. 
    We discuss the validity of this solution,
    arguing that the very interesting and peculiar properties of the solution
    are associated to the negative energy contributions coming from the terms
    quadratic in the curvature. As a matter of fact, the 10-dimensional
    configuration is singular. We extract some general lessons on attempts to
    eliminate black-hole singularities by introducing terms of higher order in
    the curvature.
\end{abstract}

\maketitle


A very well-known class of 4-dimensional extremal stringy black holes is
characterized by 4 real functions  $\mathcal{Z}_{0},\,
\mathcal{Z}_{+},\, \mathcal{Z}_{-},\, \mathcal{H}$ (which are harmonic in
3-dimensional Euclidean space $\mathbb{E}^{3}$ at zeroth order in $\alpha'$) that occur in the
metric and real scalar fields $\phi,k,\ell$ as follows: \footnote{For the sake
  of simplicity, we omit the 4 non-vanishing vector fields associated to the 4
  harmonic functions. They can be derived from the general 10-dimensional
  solution given in Ref.~\cite{Chimento:2018kop}.}

\begin{equation}
\label{eq:4chargebh}
\begin{array}{rclrcl}
ds^{2} 
& = & 
e^{2U}dt^{2}-e^{-2U}d\vec{x}^{\, 2}\, ,
& &   \\
& & & & & \\
e^{-2U} 
& = &
\sqrt{~\mathcal{Z}_{0}\, \mathcal{Z}_{+}\, \mathcal{Z}_{-}\, \mathcal{H}~}\, ,
&
e^{2\phi}
& = &
e^{2\phi_{\infty}}{\displaystyle\frac{\mathcal{Z}_{0}}{\mathcal{Z}_{-}}}\, ,
\\
& & & & & \\
\ell
& = & 
{\displaystyle
\ell_{\infty}
\left(\frac{\mathcal{Z}_{0}\mathcal{Z}_{+}\mathcal{Z}_{-}}{\mathcal{H}^{3}}\right)^{1/6}\, , 
}
&
k
& = & 
{\displaystyle
k_{\infty} 
\left(\frac{\mathcal{Z}_{+}^{2}}{\mathcal{Z}_{-}\mathcal{Z}_{0}}\right)^{1/4}\, .  }
\end{array}
\end{equation}

This 4-dimensional configuration can be obtained from a 10-dimensional
(zeroth-order in $\alpha'$) solution of the Heterotic Superstring effective
action of the form considered in Ref.~\cite{Chimento:2018kop}. They describe
fundamental strings (associated to $\mathcal{Z}_{-}$), Kaluza-Klein (KK) monopoles
(associated to $\mathcal{H}$), solitonic (NS) 5-branes (associated to
$\mathcal{Z}_{0}$) and waves traveling along the fundamental strings
(associated to $\mathcal{Z}_{+}$). These configurations are also solutions of
the STU model that arises in the compactification on a $T^{6}$ of the
10-dimensional theory \cite{Duff:1995sm,Rahmfeld:1995fm}.

In Ref.~\cite{Chimento:2018kop} we showed that $\mathcal{Z}_{-}$ and
$\mathcal{H}$ do not receive any first-order $\alpha'$ corrections. We also
showed that $\mathcal{Z}_{0}$ does have corrections, but all of them can be
eliminated by choosing appropriate SU$(2)$ instantons.  These solutions are
called ``symmetric'', and in absence of other charges they have been argued to
be exact solutions to all orders in $\alpha'$. Finally, $\mathcal{Z}_{+}$ also
has first-order $\alpha'$ corrections of the form \footnote{
  Observe that these corrections are non-vanishing even for trivial
  $\mathcal{Z}_{0}$ and $\mathcal{H}$, in which case the solutions would be
  special cases of the \textit{chiral null model} discussed in
  Refs.~\cite{Behrndt:1995tr, Horowitz:1994rf}. In those references it was
  argued that those solutions describing fundamental strings and wave
  traveling along them receive no $\alpha'$ corrections and are exact to all
  orders in $\alpha'$ in some renormalization scheme which is natural for this
  model. In the context of the Heterotic Superstring and in the scheme
  associated to quartic action given in Ref.~\cite{Bergshoeff:1989de} that we
  are using, though, these corrections are expected on physical grounds
  \cite{Prester:2010cw}. Furthermore, they are expected to play an important
  role: they can resolve the singular horizon of small black holes with 2
  charges in $d=5$ dimensions \cite{Cano:2018qev} and with 2 or 3 in $d=4$, as
  we will show in a forthcoming paper in which we will study the $\alpha'$
  corrections of general 4-dimensional black holes \cite{kn:CChMORR}. It
  should also be mentioned that, being associated to the $uu$ component of the
  Einstein equation ($u$ being a null coordinate), these corrections cannot be
  seen by merely observing the curvature invariants.\label{longfoot}}

\begin{equation}
\label{eq:exp_Z+2}
\begin{aligned}
\mathcal{Z}_{+} 
&=  
 1+\frac{q_{+}}{r}\\
&+\frac{\alpha' q_{+}}{2qq_{0}} 
\frac{r^{2}+r(q_{0}+q_{-}+q)+q q_{0}+q q_{-}+q_{0}q_{-}}{(r+q)(r+q_{0})(r+q_{-})} \\
&+\mathcal{O}(\alpha'^{2})\, ,
\end{aligned}
\end{equation}

\noindent
that cannot be cancelled by the mechanism mentioned above.

Due to the structure of the $T$-tensors, it can be argued as in
Ref.~\cite{Cano:2018qev} that the symmetric solution with any
number of charges and with just the above first-order $\alpha'$ correction of
$\mathcal{Z}_{+}$, could also be an exact solution to all orders in $\alpha'$,
or at least that the higher-order corrections would be much smaller. In other
words, the $\mathcal{O}(\alpha'^{2})$ terms could be neglected for all
purposes. It is interesting to investigate if these corrected and probably
exact solutions satisfy some of the properties that are expected to occur in a
UV complete theory, and in particular, the resolution of singularities.

As mentioned in Footnote~\footnotemark[2]{\ref{longfoot}}, we have already studied how $\alpha'$
corrections can resolve the singular horizon of small black holes (with 2 or 3
charges) as in the classical example of Ref.~\cite{Dabholkar:2004dq}, yielding
a smooth horizon with non-vanishing area, though some divergencies persist in
the KK scalars. In this paper we are going to study a particularly interesting
set of singular solutions that have 4 non-vanishing charges: massless black
holes \cite{Behrndt:1995tr,Kallosh:1995yz} (referred to as massless
\textit{quadruholes} in Ref.~\cite{Ortin:1996nd}). Research on massless black
holes was originally motivated by Strominger's description of the conifold
transition in Ref.~\cite{Strominger:1995cz}. Although his description was
based on type~II string theory and black holes with Ramond-Ramond charge, the
solutions may be related by duality and the metrics are indeed identical.


The massless quadruholes are a particular case of the solutions
Eq.~(\ref{eq:4chargebh}). They correspond to the choice 

\begin{equation}\label{eq:charges}
q_{0}=q_{-}=-q=-q_{+}=Q\geq 0\, ,
\end{equation}

\noindent
which can be achieved if the string coupling constant $g_s$ and the radii of the
compactification circles at infinity satisfy
\begin{equation}
g_s=\sqrt{\frac{N_{S5}}{N_{F1}}}\, ,\quad \frac{R_5}{\ell_s}=\sqrt{\frac{-N_{W}}{N_{F1}}}\, ,\quad \frac{R_4}{\ell_s}=\sqrt{\frac{-N_{S5}}{N_{KK}}}\, .
\end{equation}

\noindent
Here $N_{S5}$, $N_{F1}$, $N_{W}$ and $N_{KK}$ are integer numbers associated
to the stringy objects of the ten-dimensional configuration.  The usual requirements 
$g_s<<1$, $R_{4,5}>\ell_s$ are satisfied if these numbers fulfill the hierarchy 
\begin{equation}
|N_W|>N_{F1}>>N_{S5}>|N_{KK}|\, .
\end{equation}

In the absence of $\alpha'$ corrections, the metric of the massless quadruholes 
reads

\begin{equation}\label{eq:masslessqh}
ds^{2} 
 =  
\left(1-\frac{Q^{2}}{r^{2}}\right)^{-1}dt^{2}
-\left(1-\frac{Q^{2}}{r^{2}}\right)\left(dr^{2}+r^{2}d\Omega_{(2)}^{2}\right)\, .
\end{equation}

This geometry clearly contains a naked singularity at $r=Q$, where the
curvature as well as some scalars diverge. It has some interesting properties
though, such as the fact that this solution is massless and that the dilaton
takes a constant value $e^{2\phi}=e^{2\phi_{\infty}}$. The repulsive behavior
noticed in Ref.~\cite{Kallosh:1995yz} is characteristic of timelike
singularities such as those of the Reissner-Nordstr\"om or negative-mass
Schwarzschild solutions.

Taking into account the $\alpha'$ corrections given by the general formula
Eq.~(\ref{eq:exp_Z+2}), the metric function $e^{-2U}$ reads

\begin{equation}\label{eq:e2Ualpha}
e^{-2U} 
= 
\sqrt{\left(1-\frac{Q^{2}}{r^{2}}\right)^{2}
+\frac{\alpha'}{2 Q}\left(\frac{1}{r}+\frac{Q}{r^{2}}-\frac{Q^{2}}{r^{3}}\right)}\, ,
\end{equation}

\noindent
and many interesting things start happening:

\begin{enumerate}[leftmargin=*]
\item First of all, note that now this solution has a mass

\begin{equation}
  M=\frac{\alpha'}{8 Q G_{N}^{(4)}}\, .
\end{equation}

\item The geometry \footnote{The scalars $\ell$ and $k$ diverge there. This is
    a common feature of small black hole solutions and is associated to  a
    problematic (singular) compactification from 10 (actually, from 6) to 4
    dimensions. In the original picture of Ref.~\cite{Strominger:1995cz},
    there is a cycle around which a D-brane is wrapped whose volume shrinks to
    zero. The regularity of the geometry at that point in spite of the
    singularity of the scalars suggests that something strange is happening,
    as we will discuss later.} is now regular at $r=Q$. Indeed, for
  $Q^{2}>\alpha'/8$, the solution can be extended up to $r=0$, where a smooth
  $\mathrm{AdS}_{2}\times\mathrm{S}^{2}$ near-horizon geometry arises. The
  area of the horizon is given by the $\alpha'$-independent
  expression \footnote{A calculation of the $\alpha'$ corrections to the area
    that give the entropy using Wald's formula \cite{Wald:1993nt,Iyer:1994ys}
    will be given in Ref.~\cite{kn:CChMORR}.}

\begin{equation}
  A=4\pi Q^{2}\, .
\end{equation}

\noindent
This is the standard expression, in terms of the charges, for the entropy of
an extremal 4-charge black hole up to $\alpha'$ corrections. However, the
relation between the entropy and the mass is very unconventional: $A$ grows
with $Q$ while $M$ goes to zero.

\item The most striking property of the metric above is that, if
  $Q^{2}>\tfrac{\alpha'}{8}$, which corresponds to masses
  $M<\sqrt{\frac{\alpha'}{8}}/G_{N}^{(4)}$, it does not contain any
  singularity behind the horizon. In order to extend the solution beyond
  $r=0$, let us introduce the \textit{tortoise coordinate} $r_{*}$ such that
  $dr_{*}\equiv e^{-2U}dr$. We define the ingoing Eddington-Finkelstein
  coordinate

\begin{equation}
v \equiv t+r_{*}\, ,
\end{equation}

\noindent
which is constant along ingoing null radial geodesics. In terms of $v$, the
metric reads

\begin{equation}
ds^{2} = e^{2U} dv^{2}-2 dv dr-e^{-2U}r^{2}d\Omega_{(2)}^{2}\, .
\end{equation}

\noindent
The metric is clearly regular at $r=0$, and it can be extended to $r<0$. A singularity
would appear whenever $e^{-2U}=0$, but looking at (\ref{eq:e2Ualpha}), we see that
this function is strictly positive for all values of $r$ if $Q^2>\alpha'/8$.
Hence, this spacetime contains no singularity and we can extend it up to 
$r\rightarrow -\infty$, where it describes another asymptotically flat
region.

\item Without loss of generality, we can consider the motion of a test particle in
the equatorial plane $\theta=\pi/2$.  Associated to the Killing vectors $\partial_v$
 and $\partial_{\varphi}$, there are two constants of motion $\epsilon$, $L$ which
 are given by

 


\begin{eqnarray}
\label{eq:epsilon}
\epsilon
& \equiv &
e^{2U} \dot{v}-\dot{r}\, ,
\\
& & \nonumber \\  
L & \equiv & r^{2} e^{-2U}\dot{\varphi}\, .
\end{eqnarray}

\noindent
Then, we can write the  mass-shell condition as

\begin{equation}
\label{eq:massshell}
\dot{r}^{2}+V_{\text{eff}}(r)=\epsilon^{2}\, .
\end{equation}

\noindent
where the radial effective potential for massless and massive particles
($\kappa=0, 1$ resp.) is given by

\begin{equation}
V_{\text{eff}}(r)=e^{2U}\left(\kappa+ e^{2U} \frac{L^{2}}{r^{2}}\right)\, .
\end{equation}




The qualitative behavior of the geodesics can be found by studying
this effective potential, which we have plotted for several values of 
$Q$ for timelike and null geodesics in Figure~\ref{fig:effectivepotentialL=0}.

\begin{figure}
  \begin{center}
     \includegraphics[scale=0.8]{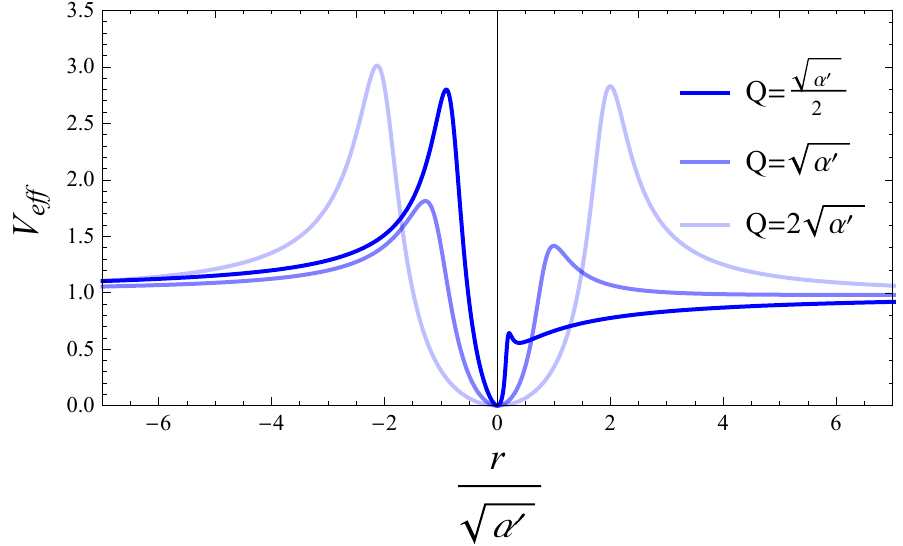}
     \includegraphics[scale=0.8]{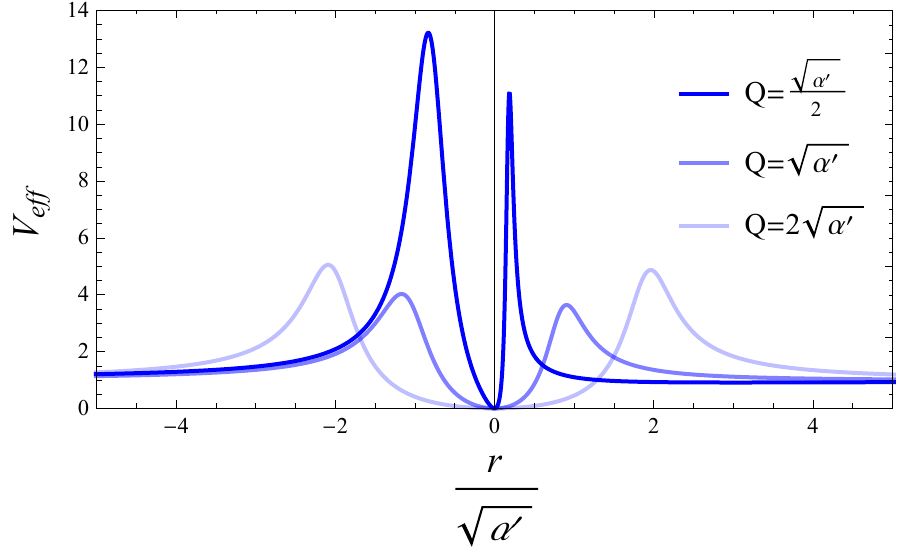}
     \includegraphics[scale=0.8]{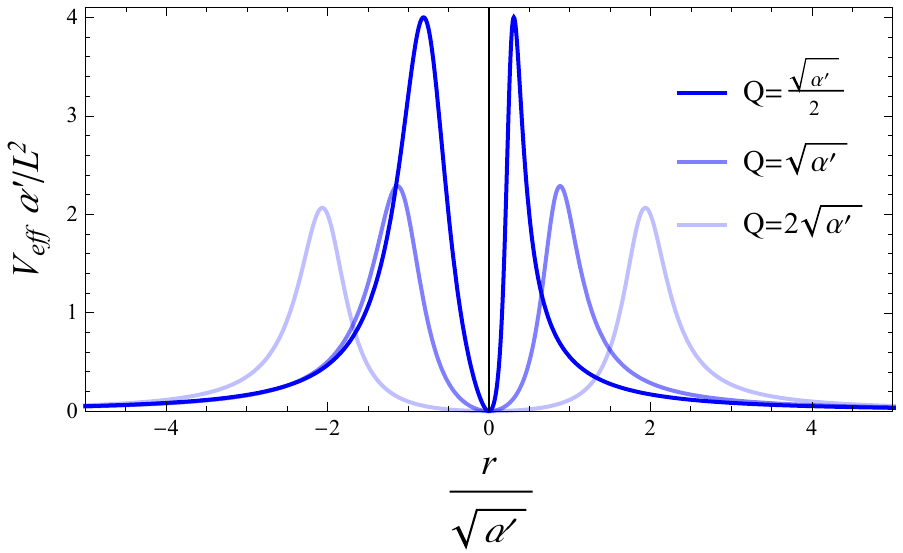}
    \caption{Effective potential for different types of geodesics and for
      several values of the charge $Q$. Top: Massive particle moving 
      along radial geodesics ($L=0$). 
      Middle: Massive particle in a non-radial geodesic with 
      $L=\alpha'$. Bottom: Massless particle 
      in a general geodesic  (for $L=0$, $V_{\text{eff}}=0$).}
      \label{fig:effectivepotentialL=0}
  \end{center}
\end{figure}

\begin{figure}
  \begin{center}
    \includegraphics[scale=0.8]{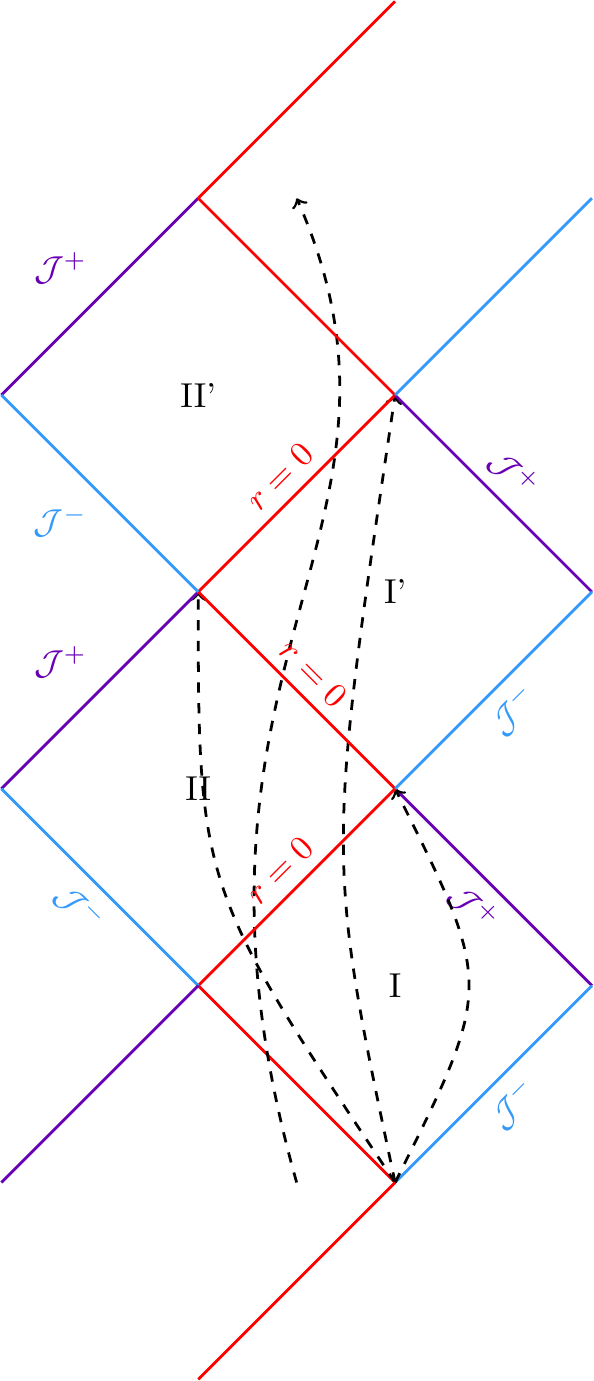}
    \caption{Penrose diagram of the $\alpha'$-corrected massless black
      holes. We have represented four kinds of possible timelike geodesics
      (described in the main text), including one corresponding to motion
      confined in the central valley of the effective potential.}
      \label{fig:Penrosediagram}
  \end{center}
\end{figure}

We see that the effective potential has a smoother form for larger values of
$Q$ \footnote{A naked singularity arises for $Q^{2}=\alpha'/8$. In the timelike
case, the height of the peaks grows as $Q/\sqrt{\alpha'/2}$ when $Q$ is large,
so the potential is actually smoother for intermediate values of $Q$.}. In all cases,
it presents two hills and a valley separating them which contains the black
hole horizon. Let us consider the geodesic of a massive particle.
Coming from large positive values of $r$, the first hill
represents a repulsive behavior which can be overcome if the particle has
enough energy. If the particle passes over the first, rightmost, hill, it will
cross the horizon at $r=0$ and it will meet the second hill, which is always
taller than the first. If the total energy is lower than the tip of the hill,
which lies at some negative value of $r$ the particle will bounce towards
larger values of $r$, but, since it is not allowed to cross the horizon (by
the very definition of event horizon) it will go into a different region of
the spacetime for which $r=0$ is a past horizon. Such a particle can reach the
infinite of the new asymptotically flat region unless some force pushes it
towards the future event horizon of that region. The same discussion can be
repeated and it is clear that an infinite number of asymptotically-flat
regions connected as in Figure~\ref{fig:Penrosediagram} exist.

If the energy of the particle is higher than the summit of the second,
leftmost, hill of the effective potential, it will cross over it towards the
$r\rightarrow -\infty$ of the type~II region in the Penrose diagram, pushed by
a repulsive force. This repulsive force is, now, associated to the negativity
of the mass of the central object as seen from the type~II regions. In some
sense, it can be said that the mass of the $\alpha'$ corrected solution
remains zero because it has opposite values in contiguous type~I and type~II
regions of the Penrose diagram.

\item The effective potential for null geodesics, plotted in
  Figure~\ref{fig:effectivepotentialL=0}, has two hills of the same
  height: if a light ray has small enough impact parameter $L/\epsilon$,
   then it always goes from type~I to type~II regions in the Penrose diagram. 
   The maximums of the potential are exactly
  \begin{equation}
  V_{\text{eff}}^{\text{max}}=\frac{16 L^2 Q^2}{\alpha'(8Q^2-\alpha')}\, .
  \end{equation}

\item In the central valleys of the effective potentials it is possible to
  have geodesics that never reach infinity and are confined between the
  hills. The particles cross the horizons (future and past) of different
  regions an infinite number of times and for an infinite number of regions as
  shown in the Penrose diagram.

\item Finally, observe that, in the timelike case the effective potential has
  another minimum in the right-hand side of the diagram (type~I region) that
  becomes shallower the larger $Q$ is. This happens even for $L=0$ (radial
  motion), which means that there can be massive particles with purely radial
  motion confined between two radii.

\end{enumerate}

Summarizing, we have found that $\alpha'$ corrections transform the 
singular massless black holes (\ref{eq:masslessqh}) into a geodesically 
complete spacetime which represents a regular black hole with no singularity. 
We focused on the particular example of the massless black holes for
convenience, but the same result is found for more general values of the
charges, provided that their signs are chosen as in (\ref{eq:charges}).

Being extremal and supersymmetric, these solutions might evade 
some of the stability issues related to regular black holes, as the ones
reported in \cite{Carballo-Rubio:2018pmi}. In particular, we note
that our black holes have a different structure as compared to 
the models analyzed there: they do not contain a de Sitter core, and the
instability associated to it does not directly apply.


Although we have argued that the $\alpha'$-corrected solutions may receive no
further corrections, it is not clear to us how seriously they should be taken
from the string theory point of view, \footnote{Since the dilaton is constant,
  and arbitrary, loop string corrections can be made as small as we want.}
because they are singular in $d=10$ dimensions. The singularity is to be
expected since some compactification circles diverge (shrink to zero radii in 
the dual theory) at given values of $r$. This pathology, on the other hand, 
may be interpreted as a sign of the relation between these solutions and 
the massless black holes of Ref.~\cite{Strominger:1995cz}.\\
\indent
A feature of these solutions that may also be considered as another sign of
this relation is the two-sided structure of the solution, which exhibits
masses of opposite signs in contiguous type~I and type~II regions of the
Penrose diagram. Some of the states that become massless in the conifold
transition could be 2-particle states and, therefore, they should have
opposite masses.

Despite the pathological character of the ten-dimensional solution, 
it should be noted that the cancellation of the black hole singularity in 
$d=4$ is a highly non-trivial effect 
related to the precise form of the corrections in $\mathcal{Z}_+$ given in
(\ref{eq:exp_Z+2}). This function diverges with the right degree precisely
at the points where the functions $\mathcal{H}$, $\mathcal{Z}_0$ and 
$\mathcal{Z}_-$ vanish, and this is the only way in which the singularity
could be removed. Even though the compactification is singular, everything 
conspires to produce a regular four-dimensional geometry.

Finally, since these solutions are also solutions of General Relativity with
complicated couplings to matter, an explanation for their completely out of
the ordinary features must be proposed. As we mentioned, the mass of the solution
has opposite sign in type~I  and type~II regions. The presence of negative masses is
usually associated to that of naked singularities and the absence of the latter can
only be attributed to the lack of positivity of the energy in the theory that
we are considering. The terms of higher order in the curvature associated to
the $\alpha'$ corrections typically have the wrong sign compared with terms
quadratic in Yang-Mills curvatures \footnote{The particular structure of these
  terms at first order in $\alpha'$ in the Heterotic Superstring Effective
  action makes the comparison possible.}. Repulsive gravitational behavior
associated to these corrections associated to these terms has been observed,
for instance, in Ref.~\cite{Hubeny:2004ji}. 

There have been many attempts in the literature to get rid of the
singularities at the core of black holes, though the analysis is usually
restricted to finding appropriate regular black hole models \cite{Frolov:2016pav,Bronnikov:2006fu}.
The theories that achieve that goal usually introduce higher-derivative
terms in the curvature \cite{Modesto:2010uh,Olmo:2015axa,Bejarano:2017fgz,Menchon:2017qed} 
or in the matter fields \cite{AyonBeato:1998ub,AyonBeato:1999rg} which may (or may not) be 
associated to quantum corrections of a theory of quantum gravity 
such as string theory. Effectively, many of those terms may introduce 
negative energy in the theory in a more or less
consistent way (nobody really knows) that it is ultimately responsible for the
removal or softening of the singularities. We believe that this aspect of the
higher-derivative terms deserves to be understood in depth if these theories are to
be considered internally consistent.

\begin{acknowledgments}
\section*{Acknowledgments}

The authors would like to thank Patrick Meessen and Pedro F.~Ram\'{\i}rez for
many useful conversations.  This work has been supported in part by the
MINECO/FEDER, UE grant FPA2015-66793-P and by the Spanish Research Agency
(Agencia Estatal de Investigaci\'on) through the grant IFT Centro de
Excelencia Severo Ochoa SEV-2016-0597.  The work of PAC was funded by
Fundaci\'on la Caixa through a ``la Caixa - Severo Ochoa'' international
pre-doctoral grant. The work of AR was
supported by a ``Centro de Excelencia Internacional UAM/CSIC'' pre-doctoral grant and by a ``Residencia
de Estudiantes'' scholarship. TO wishes to thank M.M.~Fern\'andez for her permanent
support.

\end{acknowledgments}



\bibliography{mybib}{}

\begin{thebibliography}{32}%
\makeatletter
\providecommand \@ifxundefined [1]{%
 \@ifx{#1\undefined}
}%
\providecommand \@ifnum [1]{%
 \ifnum #1\expandafter \@firstoftwo
 \else \expandafter \@secondoftwo
 \fi
}%
\providecommand \@ifx [1]{%
 \ifx #1\expandafter \@firstoftwo
 \else \expandafter \@secondoftwo
 \fi
}%
\providecommand \natexlab [1]{#1}%
\providecommand \enquote  [1]{``#1''}%
\providecommand \bibnamefont  [1]{#1}%
\providecommand \bibfnamefont [1]{#1}%
\providecommand \citenamefont [1]{#1}%
\providecommand \href@noop [0]{\@secondoftwo}%
\providecommand \href [0]{\begingroup \@sanitize@url \@href}%
\providecommand \@href[1]{\@@startlink{#1}\@@href}%
\providecommand \@@href[1]{\endgroup#1\@@endlink}%
\providecommand \@sanitize@url [0]{\catcode `\\12\catcode `\$12\catcode
  `\&12\catcode `\#12\catcode `\^12\catcode `\_12\catcode `\%12\relax}%
\providecommand \@@startlink[1]{}%
\providecommand \@@endlink[0]{}%
\providecommand \url  [0]{\begingroup\@sanitize@url \@url }%
\providecommand \@url [1]{\endgroup\@href {#1}{\urlprefix }}%
\providecommand \urlprefix  [0]{URL }%
\providecommand \Eprint [0]{\href }%
\providecommand \doibase [0]{http://dx.doi.org/}%
\providecommand \selectlanguage [0]{\@gobble}%
\providecommand \bibinfo  [0]{\@secondoftwo}%
\providecommand \bibfield  [0]{\@secondoftwo}%
\providecommand \translation [1]{[#1]}%
\providecommand \BibitemOpen [0]{}%
\providecommand \bibitemStop [0]{}%
\providecommand \bibitemNoStop [0]{.\EOS\space}%
\providecommand \EOS [0]{\spacefactor3000\relax}%
\providecommand \BibitemShut  [1]{\csname bibitem#1\endcsname}%
\let\auto@bib@innerbib\@empty
\bibitem [{Note1()}]{Note1}%
  \BibitemOpen
  \bibinfo {note} {For the sake of simplicity, we omit the 4 non-vanishing
  vector fields associated to the 4 harmonic functions. They can be derived
  from the general 10-dimensional solution given in Ref.~\cite
  {Chimento:2018kop}.}\BibitemShut {Stop}%
\bibitem [{\citenamefont {Chimento}\ \emph {et~al.}(2018)\citenamefont
  {Chimento}, \citenamefont {Meessen}, \citenamefont {Ortin}, \citenamefont
  {Ramirez},\ and\ \citenamefont {Ruiperez}}]{Chimento:2018kop}%
  \BibitemOpen
  \bibfield  {author} {\bibinfo {author} {\bibfnamefont {S.}~\bibnamefont
  {Chimento}}, \bibinfo {author} {\bibfnamefont {P.}~\bibnamefont {Meessen}},
  \bibinfo {author} {\bibfnamefont {T.}~\bibnamefont {Ortin}}, \bibinfo
  {author} {\bibfnamefont {P.~F.}\ \bibnamefont {Ramirez}}, \ and\ \bibinfo
  {author} {\bibfnamefont {A.}~\bibnamefont {Ruiperez}},\ }\href@noop {} {\
  (\bibinfo {year} {2018})},\ \Eprint {http://arxiv.org/abs/1803.04463}
  {arXiv:1803.04463 [hep-th]} \BibitemShut {NoStop}%
\bibitem [{\citenamefont {Duff}\ \emph {et~al.}(1996)\citenamefont {Duff},
  \citenamefont {Liu},\ and\ \citenamefont {Rahmfeld}}]{Duff:1995sm}%
  \BibitemOpen
  \bibfield  {author} {\bibinfo {author} {\bibfnamefont {M.~J.}\ \bibnamefont
  {Duff}}, \bibinfo {author} {\bibfnamefont {J.~T.}\ \bibnamefont {Liu}}, \
  and\ \bibinfo {author} {\bibfnamefont {J.}~\bibnamefont {Rahmfeld}},\ }\href
  {\doibase 10.1016/0550-3213(95)00555-2} {\bibfield  {journal} {\bibinfo
  {journal} {Nucl. Phys.}\ }\textbf {\bibinfo {volume} {B459}},\ \bibinfo
  {pages} {125} (\bibinfo {year} {1996})},\ \Eprint
  {http://arxiv.org/abs/hep-th/9508094} {arXiv:hep-th/9508094 [hep-th]}
  \BibitemShut {NoStop}%
\bibitem [{\citenamefont {Rahmfeld}(1996)}]{Rahmfeld:1995fm}%
  \BibitemOpen
  \bibfield  {author} {\bibinfo {author} {\bibfnamefont {J.}~\bibnamefont
  {Rahmfeld}},\ }\href {\doibase 10.1016/0370-2693(96)00063-9} {\bibfield
  {journal} {\bibinfo  {journal} {Phys. Lett.}\ }\textbf {\bibinfo {volume}
  {B372}},\ \bibinfo {pages} {198} (\bibinfo {year} {1996})},\ \Eprint
  {http://arxiv.org/abs/hep-th/9512089} {arXiv:hep-th/9512089 [hep-th]}
  \BibitemShut {NoStop}%
\bibitem [{Note2()}]{Note2}%
  \BibitemOpen
  \bibinfo {note} {Observe that these corrections are non-vanishing even for
  trivial $\protect \mathcal {Z}_{0}$ and $\protect \mathcal {H}$, in which
  case the solutions would be special cases of the \protect \textit {chiral
  null model} discussed in Refs.~\cite {Behrndt:1995tr, Horowitz:1994rf}. In
  those references it was argued that those solutions describing fundamental
  strings and wave traveling along them receive no $\alpha '$ corrections and
  are exact to all orders in $\alpha '$ in some renormalization scheme which is
  natural for this model. In the context of the Heterotic Superstring and in
  the scheme associated to quartic action given in Ref.~\cite
  {Bergshoeff:1989de} that we are using, though, these corrections are expected
  on physical grounds \cite {Prester:2010cw}. Furthermore, they are expected to
  play an important role: they can resolve the singular horizon of small black
  holes with 2 charges in $d=5$ dimensions \cite {Cano:2018qev} and with 2 or 3
  in $d=4$, as we will show in a forthcoming paper in which we will study the
  $\alpha '$ corrections of general 4-dimensional black holes \cite
  {kn:CChMORR}. It should also be mentioned that, being associated to the $uu$
  component of the Einstein equation ($u$ being a null coordinate), these
  corrections cannot be seen by merely observing the curvature
  invariants.\label {longfoot}}\BibitemShut {NoStop}%
\bibitem [{\citenamefont {Cano}\ \emph {et~al.}(2018)\citenamefont {Cano},
  \citenamefont {Meessen}, \citenamefont {Ort\'in},\ and\ \citenamefont
  {Ram\'irez}}]{Cano:2018qev}%
  \BibitemOpen
  \bibfield  {author} {\bibinfo {author} {\bibfnamefont {P.~A.}\ \bibnamefont
  {Cano}}, \bibinfo {author} {\bibfnamefont {P.}~\bibnamefont {Meessen}},
  \bibinfo {author} {\bibfnamefont {T.}~\bibnamefont {Ort\'in}}, \ and\
  \bibinfo {author} {\bibfnamefont {P.~F.}\ \bibnamefont {Ram\'irez}},\ }\href
  {\doibase 10.1007/JHEP05(2018)110} {\bibfield  {journal} {\bibinfo  {journal}
  {JHEP}\ }\textbf {\bibinfo {volume} {05}},\ \bibinfo {pages} {110} (\bibinfo
  {year} {2018})},\ \Eprint {http://arxiv.org/abs/1803.01919} {arXiv:1803.01919
  [hep-th]} \BibitemShut {NoStop}%
\bibitem [{\citenamefont {Dabholkar}\ \emph {et~al.}(2004)\citenamefont
  {Dabholkar}, \citenamefont {Kallosh},\ and\ \citenamefont
  {Maloney}}]{Dabholkar:2004dq}%
  \BibitemOpen
  \bibfield  {author} {\bibinfo {author} {\bibfnamefont {A.}~\bibnamefont
  {Dabholkar}}, \bibinfo {author} {\bibfnamefont {R.}~\bibnamefont {Kallosh}},
  \ and\ \bibinfo {author} {\bibfnamefont {A.}~\bibnamefont {Maloney}},\ }\href
  {\doibase 10.1088/1126-6708/2004/12/059} {\bibfield  {journal} {\bibinfo
  {journal} {JHEP}\ }\textbf {\bibinfo {volume} {12}},\ \bibinfo {pages} {059}
  (\bibinfo {year} {2004})},\ \Eprint {http://arxiv.org/abs/hep-th/0410076}
  {arXiv:hep-th/0410076 [hep-th]} \BibitemShut {NoStop}%
\bibitem [{\citenamefont {Behrndt}(1995)}]{Behrndt:1995tr}%
  \BibitemOpen
  \bibfield  {author} {\bibinfo {author} {\bibfnamefont {K.}~\bibnamefont
  {Behrndt}},\ }\href {\doibase 10.1016/0550-3213(95)00506-N} {\bibfield
  {journal} {\bibinfo  {journal} {Nucl. Phys.}\ }\textbf {\bibinfo {volume}
  {B455}},\ \bibinfo {pages} {188} (\bibinfo {year} {1995})},\ \Eprint
  {http://arxiv.org/abs/hep-th/9506106} {arXiv:hep-th/9506106 [hep-th]}
  \BibitemShut {NoStop}%
\bibitem [{\citenamefont {Kallosh}\ and\ \citenamefont
  {Linde}(1995)}]{Kallosh:1995yz}%
  \BibitemOpen
  \bibfield  {author} {\bibinfo {author} {\bibfnamefont {R.}~\bibnamefont
  {Kallosh}}\ and\ \bibinfo {author} {\bibfnamefont {A.~D.}\ \bibnamefont
  {Linde}},\ }\href {\doibase 10.1103/PhysRevD.52.7137} {\bibfield  {journal}
  {\bibinfo  {journal} {Phys. Rev.}\ }\textbf {\bibinfo {volume} {D52}},\
  \bibinfo {pages} {7137} (\bibinfo {year} {1995})},\ \Eprint
  {http://arxiv.org/abs/hep-th/9507022} {arXiv:hep-th/9507022 [hep-th]}
  \BibitemShut {NoStop}%
\bibitem [{\citenamefont {Ortin}(1996)}]{Ortin:1996nd}%
  \BibitemOpen
  \bibfield  {author} {\bibinfo {author} {\bibfnamefont {T.}~\bibnamefont
  {Ortin}},\ }\href {\doibase 10.1103/PhysRevLett.76.3890} {\bibfield
  {journal} {\bibinfo  {journal} {Phys. Rev. Lett.}\ }\textbf {\bibinfo
  {volume} {76}},\ \bibinfo {pages} {3890} (\bibinfo {year} {1996})},\ \Eprint
  {http://arxiv.org/abs/hep-th/9602067} {arXiv:hep-th/9602067 [hep-th]}
  \BibitemShut {NoStop}%
\bibitem [{\citenamefont {Strominger}(1995)}]{Strominger:1995cz}%
  \BibitemOpen
  \bibfield  {author} {\bibinfo {author} {\bibfnamefont {A.}~\bibnamefont
  {Strominger}},\ }\href {\doibase 10.1016/0550-3213(95)00287-3} {\bibfield
  {journal} {\bibinfo  {journal} {Nucl. Phys.}\ }\textbf {\bibinfo {volume}
  {B451}},\ \bibinfo {pages} {96} (\bibinfo {year} {1995})},\ \Eprint
  {http://arxiv.org/abs/hep-th/9504090} {arXiv:hep-th/9504090 [hep-th]}
  \BibitemShut {NoStop}%
\bibitem [{Note3()}]{Note3}%
  \BibitemOpen
  \bibinfo {note} {The scalars $\ell $ and $k$ diverge there. This is a common
  feature of small black hole solutions and is associated to a problematic
  (singular) compactification from 10 (actually, from 6) to 4 dimensions. In
  the original picture of Ref.~\cite {Strominger:1995cz}, there is a cycle
  around which a D-brane is wrapped whose volume shrinks to zero. The
  regularity of the geometry at that point in spite of the singularity of the
  scalars suggests that something strange is happening, as we will discuss
  later.}\BibitemShut {Stop}%
\bibitem [{Note4()}]{Note4}%
  \BibitemOpen
  \bibinfo {note} {A calculation of the $\alpha '$ corrections to the area that
  give the entropy using Wald's formula \cite {Wald:1993nt,Iyer:1994ys} will be
  given in Ref.~\cite {kn:CChMORR}.}\BibitemShut {Stop}%
\bibitem [{Note5()}]{Note5}%
  \BibitemOpen
  \bibinfo {note} {A naked singularity arises for $Q^{2}=\alpha '/8$. In the
  timelike case, the height of the peaks grows as $Q/\protect \sqrt {\alpha
  '/2}$ when $Q$ is large, so the potential is actually smoother for
  intermediate values of $Q$.}\BibitemShut {Stop}%
\bibitem [{\citenamefont {Carballo-Rubio}\ \emph {et~al.}(2018)\citenamefont
  {Carballo-Rubio}, \citenamefont {Di~Filippo}, \citenamefont {Liberati},
  \citenamefont {Pacilio},\ and\ \citenamefont
  {Visser}}]{Carballo-Rubio:2018pmi}%
  \BibitemOpen
  \bibfield  {author} {\bibinfo {author} {\bibfnamefont {R.}~\bibnamefont
  {Carballo-Rubio}}, \bibinfo {author} {\bibfnamefont {F.}~\bibnamefont
  {Di~Filippo}}, \bibinfo {author} {\bibfnamefont {S.}~\bibnamefont
  {Liberati}}, \bibinfo {author} {\bibfnamefont {C.}~\bibnamefont {Pacilio}}, \
  and\ \bibinfo {author} {\bibfnamefont {M.}~\bibnamefont {Visser}},\ }\href
  {\doibase 10.1007/JHEP07(2018)023} {\bibfield  {journal} {\bibinfo  {journal}
  {JHEP}\ }\textbf {\bibinfo {volume} {07}},\ \bibinfo {pages} {023} (\bibinfo
  {year} {2018})},\ \Eprint {http://arxiv.org/abs/1805.02675} {arXiv:1805.02675
  [gr-qc]} \BibitemShut {NoStop}%
\bibitem [{Note6()}]{Note6}%
  \BibitemOpen
  \bibinfo {note} {Since the dilaton is constant, and arbitrary, loop string
  corrections can be made as small as we want.}\BibitemShut {Stop}%
\bibitem [{Note7()}]{Note7}%
  \BibitemOpen
  \bibinfo {note} {The particular structure of these terms at first order in
  $\alpha '$ in the Heterotic Superstring Effective action makes the comparison
  possible.}\BibitemShut {Stop}%
\bibitem [{\citenamefont {Hubeny}\ \emph {et~al.}(2005)\citenamefont {Hubeny},
  \citenamefont {Maloney},\ and\ \citenamefont {Rangamani}}]{Hubeny:2004ji}%
  \BibitemOpen
  \bibfield  {author} {\bibinfo {author} {\bibfnamefont {V.}~\bibnamefont
  {Hubeny}}, \bibinfo {author} {\bibfnamefont {A.}~\bibnamefont {Maloney}}, \
  and\ \bibinfo {author} {\bibfnamefont {M.}~\bibnamefont {Rangamani}},\ }\href
  {\doibase 10.1088/1126-6708/2005/05/035} {\bibfield  {journal} {\bibinfo
  {journal} {JHEP}\ }\textbf {\bibinfo {volume} {05}},\ \bibinfo {pages} {035}
  (\bibinfo {year} {2005})},\ \Eprint {http://arxiv.org/abs/hep-th/0411272}
  {arXiv:hep-th/0411272 [hep-th]} \BibitemShut {NoStop}%
\bibitem [{\citenamefont {Frolov}(2016)}]{Frolov:2016pav}%
  \BibitemOpen
  \bibfield  {author} {\bibinfo {author} {\bibfnamefont {V.~P.}\ \bibnamefont
  {Frolov}},\ }\href {\doibase 10.1103/PhysRevD.94.104056} {\bibfield
  {journal} {\bibinfo  {journal} {Phys. Rev.}\ }\textbf {\bibinfo {volume}
  {D94}},\ \bibinfo {pages} {104056} (\bibinfo {year} {2016})},\ \Eprint
  {http://arxiv.org/abs/1609.01758} {arXiv:1609.01758 [gr-qc]} \BibitemShut
  {NoStop}%
\bibitem [{\citenamefont {Bronnikov}\ \emph {et~al.}(2007)\citenamefont
  {Bronnikov}, \citenamefont {Melnikov},\ and\ \citenamefont
  {Dehnen}}]{Bronnikov:2006fu}%
  \BibitemOpen
  \bibfield  {author} {\bibinfo {author} {\bibfnamefont {K.~A.}\ \bibnamefont
  {Bronnikov}}, \bibinfo {author} {\bibfnamefont {V.~N.}\ \bibnamefont
  {Melnikov}}, \ and\ \bibinfo {author} {\bibfnamefont {H.}~\bibnamefont
  {Dehnen}},\ }\href {\doibase 10.1007/s10714-007-0430-6} {\bibfield  {journal}
  {\bibinfo  {journal} {Gen. Rel. Grav.}\ }\textbf {\bibinfo {volume} {39}},\
  \bibinfo {pages} {973} (\bibinfo {year} {2007})},\ \Eprint
  {http://arxiv.org/abs/gr-qc/0611022} {arXiv:gr-qc/0611022 [gr-qc]}
  \BibitemShut {NoStop}%
\bibitem [{\citenamefont {Modesto}\ \emph {et~al.}(2011)\citenamefont
  {Modesto}, \citenamefont {Moffat},\ and\ \citenamefont
  {Nicolini}}]{Modesto:2010uh}%
  \BibitemOpen
  \bibfield  {author} {\bibinfo {author} {\bibfnamefont {L.}~\bibnamefont
  {Modesto}}, \bibinfo {author} {\bibfnamefont {J.~W.}\ \bibnamefont {Moffat}},
  \ and\ \bibinfo {author} {\bibfnamefont {P.}~\bibnamefont {Nicolini}},\
  }\href {\doibase 10.1016/j.physletb.2010.11.046} {\bibfield  {journal}
  {\bibinfo  {journal} {Phys. Lett.}\ }\textbf {\bibinfo {volume} {B695}},\
  \bibinfo {pages} {397} (\bibinfo {year} {2011})},\ \Eprint
  {http://arxiv.org/abs/1010.0680} {arXiv:1010.0680 [gr-qc]} \BibitemShut
  {NoStop}%
\bibitem [{\citenamefont {Olmo}\ and\ \citenamefont
  {Rubiera-Garcia}(2015)}]{Olmo:2015axa}%
  \BibitemOpen
  \bibfield  {author} {\bibinfo {author} {\bibfnamefont {G.~J.}\ \bibnamefont
  {Olmo}}\ and\ \bibinfo {author} {\bibfnamefont {D.}~\bibnamefont
  {Rubiera-Garcia}},\ }\href {\doibase 10.3390/universe1020173} {\bibfield
  {journal} {\bibinfo  {journal} {Universe}\ }\textbf {\bibinfo {volume} {1}},\
  \bibinfo {pages} {173} (\bibinfo {year} {2015})},\ \Eprint
  {http://arxiv.org/abs/1509.02430} {arXiv:1509.02430 [hep-th]} \BibitemShut
  {NoStop}%
\bibitem [{\citenamefont {Bejarano}\ \emph {et~al.}(2017)\citenamefont
  {Bejarano}, \citenamefont {Olmo},\ and\ \citenamefont
  {Rubiera-Garcia}}]{Bejarano:2017fgz}%
  \BibitemOpen
  \bibfield  {author} {\bibinfo {author} {\bibfnamefont {C.}~\bibnamefont
  {Bejarano}}, \bibinfo {author} {\bibfnamefont {G.~J.}\ \bibnamefont {Olmo}},
  \ and\ \bibinfo {author} {\bibfnamefont {D.}~\bibnamefont {Rubiera-Garcia}},\
  }\href {\doibase 10.1103/PhysRevD.95.064043} {\bibfield  {journal} {\bibinfo
  {journal} {Phys. Rev.}\ }\textbf {\bibinfo {volume} {D95}},\ \bibinfo {pages}
  {064043} (\bibinfo {year} {2017})},\ \Eprint
  {http://arxiv.org/abs/1702.01292} {arXiv:1702.01292 [hep-th]} \BibitemShut
  {NoStop}%
\bibitem [{\citenamefont {Menchon}\ \emph {et~al.}(2017)\citenamefont
  {Menchon}, \citenamefont {Olmo},\ and\ \citenamefont
  {Rubiera-Garcia}}]{Menchon:2017qed}%
  \BibitemOpen
  \bibfield  {author} {\bibinfo {author} {\bibfnamefont {C.}~\bibnamefont
  {Menchon}}, \bibinfo {author} {\bibfnamefont {G.~J.}\ \bibnamefont {Olmo}}, \
  and\ \bibinfo {author} {\bibfnamefont {D.}~\bibnamefont {Rubiera-Garcia}},\
  }\href {\doibase 10.1103/PhysRevD.96.104028} {\bibfield  {journal} {\bibinfo
  {journal} {Phys. Rev.}\ }\textbf {\bibinfo {volume} {D96}},\ \bibinfo {pages}
  {104028} (\bibinfo {year} {2017})},\ \Eprint
  {http://arxiv.org/abs/1709.09592} {arXiv:1709.09592 [gr-qc]} \BibitemShut
  {NoStop}%
\bibitem [{\citenamefont {Ayon-Beato}\ and\ \citenamefont
  {Garcia}(1998)}]{AyonBeato:1998ub}%
  \BibitemOpen
  \bibfield  {author} {\bibinfo {author} {\bibfnamefont {E.}~\bibnamefont
  {Ayon-Beato}}\ and\ \bibinfo {author} {\bibfnamefont {A.}~\bibnamefont
  {Garcia}},\ }\href {\doibase 10.1103/PhysRevLett.80.5056} {\bibfield
  {journal} {\bibinfo  {journal} {Phys. Rev. Lett.}\ }\textbf {\bibinfo
  {volume} {80}},\ \bibinfo {pages} {5056} (\bibinfo {year} {1998})},\ \Eprint
  {http://arxiv.org/abs/gr-qc/9911046} {arXiv:gr-qc/9911046 [gr-qc]}
  \BibitemShut {NoStop}%
\bibitem [{\citenamefont {Ayon-Beato}\ and\ \citenamefont
  {Garcia}(1999)}]{AyonBeato:1999rg}%
  \BibitemOpen
  \bibfield  {author} {\bibinfo {author} {\bibfnamefont {E.}~\bibnamefont
  {Ayon-Beato}}\ and\ \bibinfo {author} {\bibfnamefont {A.}~\bibnamefont
  {Garcia}},\ }\href {\doibase 10.1016/S0370-2693(99)01038-2} {\bibfield
  {journal} {\bibinfo  {journal} {Phys. Lett.}\ }\textbf {\bibinfo {volume}
  {B464}},\ \bibinfo {pages} {25} (\bibinfo {year} {1999})},\ \Eprint
  {http://arxiv.org/abs/hep-th/9911174} {arXiv:hep-th/9911174 [hep-th]}
  \BibitemShut {NoStop}%
\bibitem [{\citenamefont {Horowitz}\ and\ \citenamefont
  {Tseytlin}(1995)}]{Horowitz:1994rf}%
  \BibitemOpen
  \bibfield  {author} {\bibinfo {author} {\bibfnamefont {G.~T.}\ \bibnamefont
  {Horowitz}}\ and\ \bibinfo {author} {\bibfnamefont {A.~A.}\ \bibnamefont
  {Tseytlin}},\ }\href {\doibase 10.1103/PhysRevD.51.2896} {\bibfield
  {journal} {\bibinfo  {journal} {Phys. Rev.}\ }\textbf {\bibinfo {volume}
  {D51}},\ \bibinfo {pages} {2896} (\bibinfo {year} {1995})},\ \Eprint
  {http://arxiv.org/abs/hep-th/9409021} {arXiv:hep-th/9409021 [hep-th]}
  \BibitemShut {NoStop}%
\bibitem [{\citenamefont {Bergshoeff}\ and\ \citenamefont
  {de~Roo}(1989)}]{Bergshoeff:1989de}%
  \BibitemOpen
  \bibfield  {author} {\bibinfo {author} {\bibfnamefont {E.~A.}\ \bibnamefont
  {Bergshoeff}}\ and\ \bibinfo {author} {\bibfnamefont {M.}~\bibnamefont
  {de~Roo}},\ }\href {\doibase 10.1016/0550-3213(89)90336-2} {\bibfield
  {journal} {\bibinfo  {journal} {Nucl. Phys.}\ }\textbf {\bibinfo {volume}
  {B328}},\ \bibinfo {pages} {439} (\bibinfo {year} {1989})}\BibitemShut
  {NoStop}%
\bibitem [{\citenamefont {Dominis~Prester}(2010)}]{Prester:2010cw}%
  \BibitemOpen
  \bibfield  {author} {\bibinfo {author} {\bibfnamefont {P.}~\bibnamefont
  {Dominis~Prester}}\ }(\bibinfo {year} {2010})\ \Eprint
  {http://arxiv.org/abs/1001.1452} {arXiv:1001.1452 [hep-th]} \BibitemShut
  {NoStop}%
\bibitem [{\citenamefont {Cano}\ \emph {et~al.}(pear)\citenamefont {Cano},
  \citenamefont {Chimento}, \citenamefont {Meessen}, \citenamefont {Ortin},
  \citenamefont {Ramirez},\ and\ \citenamefont {Ruiperez}}]{kn:CChMORR}%
  \BibitemOpen
  \bibfield  {author} {\bibinfo {author} {\bibfnamefont {P.~A.}\ \bibnamefont
  {Cano}}, \bibinfo {author} {\bibfnamefont {S.}~\bibnamefont {Chimento}},
  \bibinfo {author} {\bibfnamefont {P.}~\bibnamefont {Meessen}}, \bibinfo
  {author} {\bibfnamefont {T.}~\bibnamefont {Ortin}}, \bibinfo {author}
  {\bibfnamefont {P.~F.}\ \bibnamefont {Ramirez}}, \ and\ \bibinfo {author}
  {\bibfnamefont {A.}~\bibnamefont {Ruiperez}},\ }\href@noop {} {\  (\bibinfo
  {year} {2018, \textit{to appear}})}\BibitemShut {NoStop}%
\bibitem [{\citenamefont {Wald}(1993)}]{Wald:1993nt}%
  \BibitemOpen
  \bibfield  {author} {\bibinfo {author} {\bibfnamefont {R.~M.}\ \bibnamefont
  {Wald}},\ }\href {\doibase 10.1103/PhysRevD.48.R3427} {\bibfield  {journal}
  {\bibinfo  {journal} {Phys. Rev.}\ }\textbf {\bibinfo {volume} {D48}},\
  \bibinfo {pages} {R3427} (\bibinfo {year} {1993})},\ \Eprint
  {http://arxiv.org/abs/gr-qc/9307038} {arXiv:gr-qc/9307038 [gr-qc]}
  \BibitemShut {NoStop}%
\bibitem [{\citenamefont {Iyer}\ and\ \citenamefont
  {Wald}(1994)}]{Iyer:1994ys}%
  \BibitemOpen
  \bibfield  {author} {\bibinfo {author} {\bibfnamefont {V.}~\bibnamefont
  {Iyer}}\ and\ \bibinfo {author} {\bibfnamefont {R.~M.}\ \bibnamefont
  {Wald}},\ }\href {\doibase 10.1103/PhysRevD.50.846} {\bibfield  {journal}
  {\bibinfo  {journal} {Phys. Rev.}\ }\textbf {\bibinfo {volume} {D50}},\
  \bibinfo {pages} {846} (\bibinfo {year} {1994})},\ \Eprint
  {http://arxiv.org/abs/gr-qc/9403028} {arXiv:gr-qc/9403028 [gr-qc]}
  \BibitemShut {NoStop}%
\end{thebibliography}%




\end{document}